\documentclass[fleqn,11pt]{wlscirep}
\usepackage[utf8]{inputenc}
\usepackage[T1]{fontenc}
\usepackage{bm}
\usepackage{setspace}
\usepackage{subfigure}

\usepackage{framed}
\title{Before WIMPs: Neutrinos and the origins of particle dark matter}

\author[1,2,*]{Jaco de Swart}

\affil[1]{Program in Science, Technology, and Society, and Center for Theoretical Physics, Massachusetts Institute of
Technology, Cambridge, MA, USA}
\affil[2]{Department of History and Philosophy of Science, and Kavli Institute for Cosmology, University of Cambridge, Cambridge, UK}

\affil[*]{e-mail: jgd39@cam.ac.uk}

\begin{abstract}
After four decades of null-results, the experimental quest to discover a dark matter particle ardently continues. While confidence in the leading candidate, the Weakly Interacting Massive Particle (WIMP), is waning, its explanatory success still shapes current experimental approaches and theoretical expectations---from underground detecting efforts in Europe, China, and the U.S. to modern theories of cosmic structure. This paper traces the origins of this particle-based paradigm to a critical predecessor: the neutrino-dominated universe. Proposed in the early 1970s as a solution to extra-galactic anomalies, neutrinos became the leading explanation of ‘missing matter’ following experimental hints of neutrino mass in 1980. Although short-lived, the model’s coherent picture of cosmic evolution and structure formation set strong conceptual and methodological standards—providing critical impetus to the field of particle cosmology. Examining this history invites renewed reflection on how such past aims and expectations continue to guide dark matter inquiry today.
\end{abstract}
\begin{document}

\flushbottom
\maketitle

\thispagestyle{empty}

After almost four decades of searching for the hypothetical\textit{ Weakly Interacting Massive Particles}, experiments have hit an awkward milestone. Three major experiments aiming to observe the long-favored WIMP dark matter candidate in the laboratory have detected hints of extraterrestrial \textit{neutrinos}.\cite{Aprile2024FirstXENONnT, PandaXCollaboration2024FirstPandaX-4T, Akerib2025SearchesExperiment} This achievement, although a great success in detector engineering, has long been dreaded by WIMP researchers: neutrinos coming from the sun, supernovae, and the atmosphere act as an unremovable “fog” that obscures WIMPs entering the detector (see Box 1). These inexorable neutrinos, some say, could well mean an end to WIMP searches as they make it impossible to conclusively detect the dark matter particles—potentially concluding the long reign of the WIMP paradigm.

Tragic as such an ending might be, neutrinos also serve to tell a different story about dark matter. There is a complex historical entanglement between dark matter and neutrinos that illuminates how the WIMP paradigm was established in the first place. In a strange irony of history, neutrinos were crucial to the initial success of the WIMP: cosmic neutrinos were the first dark matter particles, legitimized the field of particle cosmology, and inspired the first WIMP detectors. Unfolding the historical connections between neutrinos and WIMPs offers insights into the origin of the particle-based picture of dark matter and the explanatory standards that still guide dark matter research today—and will keep on reverberating even as the WIMP wanes. 

\section*{The birth of neutrino dark matter}
In the early 1970s, in the wake of the confirmation of the big bang theory, astronomers started to note something peculiar was going on with masses of galaxies. Although most galaxies seemed to exist in large clusters, measurements of galaxy velocities appeared to contradict that: the velocities of galaxies were too high for them to form stable systems. This discrepancy, first observed by astronomer Fritz Zwicky four decades earlier, implied that galaxy clusters are either exploding apart, or are stabilized by some unseen “missing mass.” The problem had become notorious by 1970. New X-ray studies, fueled by space-race technologies, showed that the most common explanation—hot intergalactic gas—existed in too small quantities to explain the issue.\cite{Gursky1971} Lacking a general framework for understanding how galaxy structures form in a big bang universe, it was unclear how astronomers were to move forward. “[I]t is by no means clear how we will interpret [this anomaly],” the rising star of the new science of big bang cosmology, James Peebles, concluded after reviewing the problem in his new 1971 textbook, “[it] may mean that the conventional picture misses an important physical effect.”\cite[pp. 79,115]{Peebles1971}

While astronomers weighed their options, Peebles’ cosmology textbook spread the word about the cluster problem beyond the domain of astronomy. In 1971, Ramanath Cowsik and Alex Szalay---two young physicists at the institutional peripheries of the field---became aware of the issue as they attempted to understand the new science of the big bang. Cowsik, a recent graduate in neutrino and cosmic ray physics from the Tata Institute of Fundamental Research in Bombay, India, picked up Peebles’ book after he started a temporary teaching position at the astronomy department of UC Berkeley. Szalay, a neutrino physics graduate student at the Eötvös Loránd University in Budapest, Hungary, was triggered to learn cosmology from Peebles’ textbook after reading a Russian paper on a possible cosmic neutrino background.\cite{Gershtein1966RestCosmology} As they learned about the cluster stability problem in the textbook, Cowsik and Szalay independently figured that a distribution of cosmic neutrinos might solve the issue.  

In 1972, Cowsik and Szalay published papers in which they used their newly acquired cosmology proficiency to solve Zwicky’s problem while putting limits on the mass of the neutrino. The neutrino---famously hypothesized by Wolfgang Pauli in 1931 and experimentally confirmed by Fredrick Reines and Clyde Cowan in 1957---was theorized to be massless.\cite{Kaiser2020QuantumWorld} However, experiments had yet to put strong bounds on the mass. Cowsik and Szalay showed how neutrinos, borne in great numbers from nuclear reactions after the Big Bang, could not be heavier than a few electronvolts---about 100.000 times lighter than an electron—otherwise they would exceed a recent measure of the mass density of the universe.\cite{Marx1972CosmologicalRemark,Cowsik1972}  If neutrinos happen to have this mass, Cowsik and Szalay realized, they would outweigh stars and galaxies—and dominate the gravitational dynamics of the universe. A cosmic distribution of these neutrinos, they calculated, would then be just right to solve the missing mass issue in galaxy clusters.\cite{Cowsik1973GravityAstrophysics,Szalay1976NeutrinoCosmology.}

%[Image 1: a. Alex Szalay in 1972, working on his first neutrino paper with a Hewlett-Packard programmable desktop on which he performed the calculations. Source: Personal collection of A. Szalay; b. Ramanath Cowsik ca 1992. Source: Personal collection of R. Cowsik.]

Cowsik and Szalay’s work received little attention. “Nobody bothered about it,” Cowsik recalled in an interview (R. Cowsik, oral history interview with J. de Swart, 24 November 2024). “[The work] was considered slightly crazy,” Szalay explained (A. Szalay, oral history interview with J. de Swart, 12 October 2019). There was little enthusiasm for their particle physics application to cosmology---a vastly unfamiliar practice to colleagues in both fields. Moreover, talk of missing mass in the universe had yet to become part and parcel of cosmological canon. The issue only started to gain notoriety after 1974, when two collaborations argued that the dynamics of galaxies at all scales---from solitary galaxies to groups and galaxy clusters---evidenced that galaxies must be enveloped by halos of invisible mass.\cite{DeSwart2017,deSwart2024FiveMatter} The observational support for the hypothesis soon grew stronger, particularly through precise measurements of the rapid rotation of galaxies. In 1979, astronomers Sandra Faber---a former summer student of galaxy rotation expert Vera Rubin---and John Gallagher amassed evidence that nailed down the case for halos of missing mass in a review paper: “we think it likely that the discovery of invisible matter will endure as one of the major conclusions of modern astronomy.”\cite[p. 182]{Faber1979} It was then that Cowsik and Szalay’s idea resurfaced.

\begin{figure}
    \centering
   \subfigure[]{ \includegraphics[width=0.655\linewidth]{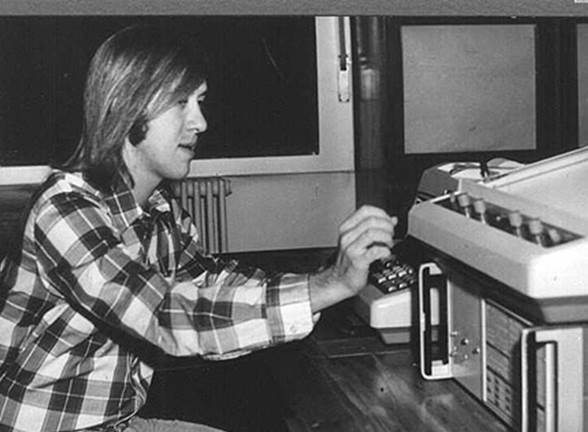}}
   \subfigure[]{ \includegraphics[width=0.309\linewidth]{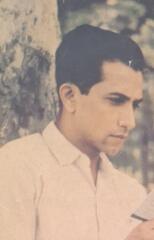}}
    \caption{a. Alex Szalay in 1972, working on his first neutrino paper with a Hewlett-Packard programmable desktop on which he performed the calculations. Source: Personal collection of A. Szalay; b. Ramanath Cowsik ca 1968, before moving to Berkeley. Source: Personal collection of R. Cowsik.}
    \label{fig:placeholder}
\end{figure}

\section*{New neutrinos in particle physics}

By the time that the issue of missing mass struck deep roots in 1979, neutrinos had taken up a different significance in physics. Neutrinos had been key to a momentous success in particle physics: the unified theory of electro-weak interactions.\cite{Galison1987,Pickering1984b} Through the experimental discovery of \textit{weak-neutral current}s at CERN in 1974, the “standard” Salam-Weinberg model of electroweak unification became widely celebrated---most notably by the 1979 Nobel Prize in Physics. The consequent detection of the J/Psi and other particles consolidated the theory of strong interactions. Neutrinos, which only interact through the weak interaction, were central to the newfangled theory and experiments. When the \textit{tau} particle was detected in 1976---the third of the lepton family, along with the electron and muon, and yet another success story of the standard model—the reality of a family of massive \textit{tau neutrinos} was immediately accepted. Together with a more spurious detection event—the tri-muon event at Fermilab\cite{Benvenuti1977CharacteristicsVertex}—the tau detection led to a surge of interest in the existence of more potential families of neutrino-like particles. How many types of neutrinos are there—and what mass do they have?

%[Image 2: The Stanford Positron Electron Asymmetric Ring (SPEAR), in December 1975. In 1975, Martin Perl and his team discovered the Tau lepton, spurring great interest in the existence of yet-undiscovered neutrino particles. Courtesy of SLAC National Accelerator Laboratory, Archives and History Office.]

The hypothesis of a yet-undiscovered heavy neutrino was the reason for a young generation of particle physicists to boldly go where few colleagues had gone before: cosmology. In tandem, in 1977, five independent groups of physicists creatively used cosmological arguments to explore the existence of hypothetical new heavy neutrinos—also termed “heavy leptons” or “massive neutral weakly interacting particles.”\cite{Hut1977,Lee1977CosmologicalMasses,Dicus1977CosmologicalLifetimes,Vysotskii1977CosmologicalLeptons,Sato1977CosmologicalNeutrinos} Like Cowsik and Szalay’s light neutrinos, these collaborations---hailing from Japan and the United States to Russia and the Netherlands---independently realized that a potential new family of neutrinos would have played a fundamental role in the thermodynamics of the early universe. If they exist, such neutrinos would have been created in primordial nuclear reactions of the big bang until their numbers “freeze out” as the universe expands and cools down. Calculating the details of this freeze-out and using observations of the maximum density of the universe, these authors put constraints on the lifetime and mass that such new neutrinos could have---and bolstered recent attempts to limit the total number of lepton families that could exist.\cite{Steigman1977CosmologicalLeptons} The conclusion? If a cosmic abundance of new, massive neutrino-like particles exists, they can only be more than several \textit{giga}-electronvolt. 

The hypothetical heavy neutrinos ushered in a new era in particle physics: cosmological arguments became aids in constraining and testing the latest ideas in particle physics.  This cosmological turn in particle physics was concurrently urged by the many new “grand unified theories” of the late 1970s---theoretical attempts to unify the strong and electro-weak forces that introduced novel hypotheticals such as proton decay and potential explanations of the cosmic asymmetry between particle and anti-particles. A generation of trained particle physicists turned into cosmologists, further prompted by curriculum changes and a plummet in particle physics funding.\cite{Kaiser2006} One of them was Steven Weinberg, author of one of the heavy neutrino papers, who published his 1977 bestseller \textit{The First Three Minutes} that explored the particle physics of the early universe.\cite{Weinberg1977} Others included Chicago heavy-weight David Schramm and his postdoc Michael Turner, who promoted the view that the universe was the “poor man's accelerator,” being able to “illuminate our knowledge of the fundamental interactions.”\cite{Turner1979CosmologyPhysics}

Spurred by the growing synthesis between particles and the cosmos, the theorized heavy neutrinos were connected to the recently corroborated missing mass problem. In 1978, two interdisciplinary collaborations of particle physicists and astronomers showed how massive neutrinos released in the big bang would take part in the evolution of the universe: they would undergo gravitational collapse together with the visible matter to form bound systems, and persist as invisible halos as visible matter dissipated energy and contracted into galaxies.
% they would trace the ordinary mass in making gravitational wells the size of clusters and form halos as the visible matter collapses into smaller objects like galaxies.
The conclusion was clear: “heavy neutrinos are an ideal material from which to form the “missing mass”,” one collaboration concluded;\cite[p.1060]{Steigman1978DynamicalNeutrinos} “[h]eavy noninteracting neutral particles [...] could not be better as stuff to constitute the dynamical missing mass” the other team wrote.\cite[p.1023]{Gunn1978SomeLepton.} Both cited Cowsik’s earlier work on the neutrino. Their idea was supported by recent calculations on primordial nucleosynthesis: the observed abundances of light elements like Helium and Lithium showed that the universe consisted only of a limited number of nucleons in gas and stars ($\Omega_{nucleon}\sim 0.04$). If so, nucleons---or \textit{baryonic matter} as the particle physicists called it---could not be the missing mass. But heavy neutrinos could.\cite{Yang1979ConstraintsNucleosynthesis.}

\section*{The neutrino-dominated universe}

The case for neutrinos as the missing matter got dramatically confirmed in the spring of 1980. During an American Physical Society meeting, Fredrick Reines of neutrino discovery fame, dramatically announced to have found evidence that neutrinos \textit{oscillate}---a result which demanded neutrinos to have mass.\cite{AIPNewsRelease1980Highlights1980, Reines1980EvidenceInstability} Even more astounding news came a month later: an experiment lead by Russian physicist Valentin Lyubimov at the Institute of Theoretical and Experimental Physics in Moscow found evidence that neutrinos had a mass of around 30 eV.\cite{Lubimov1980AnMolecule}  This was not the recently hypothesized “heavy neutrino”, but a massive \textit{electron neutrino}, exactly in the mass range explored by Cowsik and Szalay almost a decade earlier. The results “opened the flood gates,” Nobel Prize laureate James Peebles remembered (P. J. E Peebles, oral history interview with J. de Swart, 12 December 2014). If it was true, the light neutrino would make up 90\% of the universe mass budget: it would provide the missing mass in clusters and halos around galaxies just as Szalay and Cowsik had shown. But there was more. These light neutrinos would also increase the mass of the universe such that it be “flat”---a long preferred model of the Big Bang in which there was enough mass for gravity to stop the force of cosmic expansion (a result that was also demanded by the theory of inflation, introduced in 1981). Massive neutrinos even supported some grand unified theories which demanded neutrinos to have mass. 

\begin{figure}
    \centering
    \includegraphics[width=0.95\linewidth]{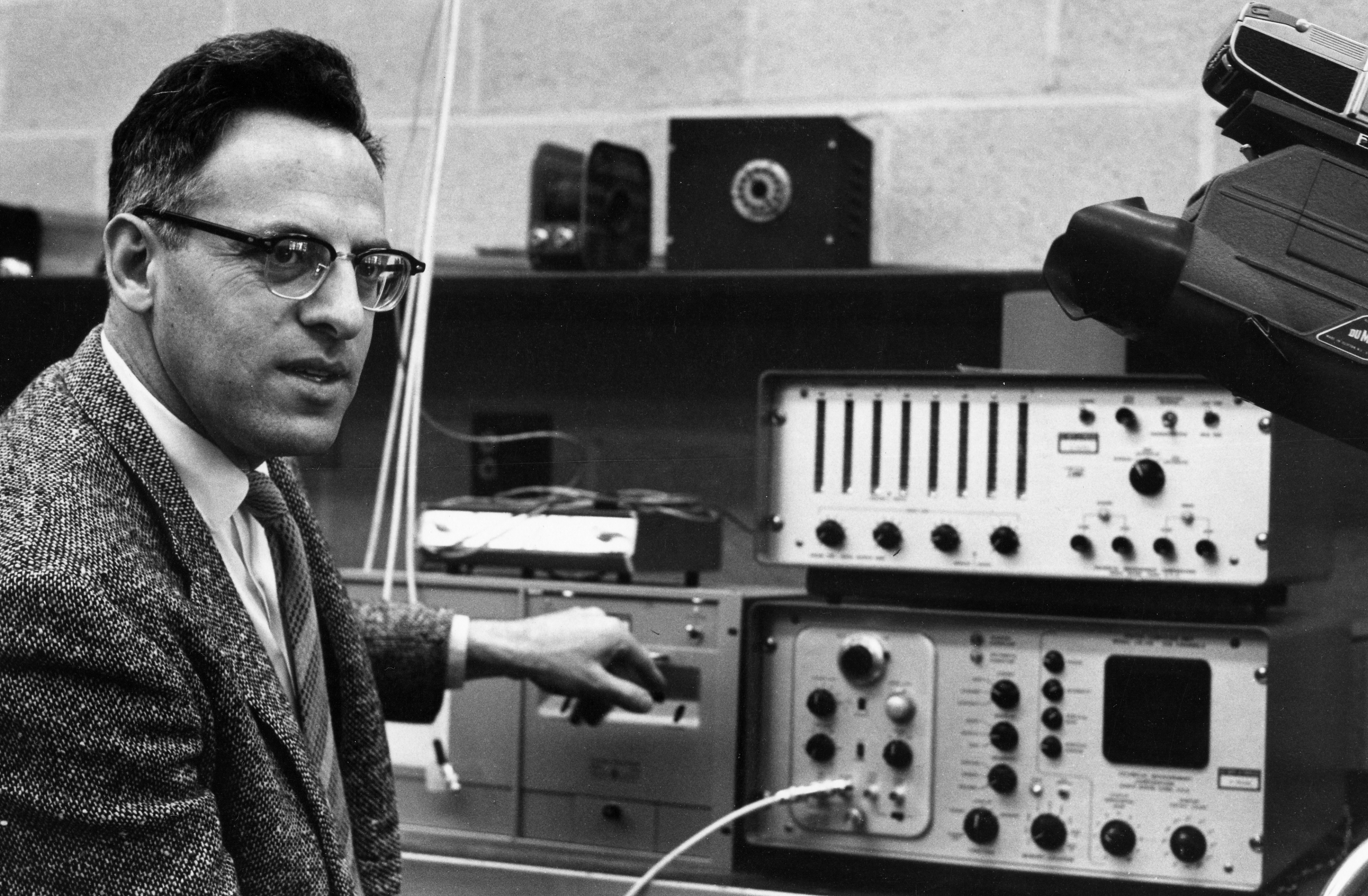}
    \caption{Fredrick Reines in an undated picture. Reines proved the existence of the neutrino together with Clyde Cowan in 1956. In 1980, he announced that he had found evidence for neutrino oscillations, implying that neutrinos have mass. Photograph by Ed Nano, courtesy of AIP Emilio Segrè Visual Archives.}
    \label{fig:placeholder}
\end{figure}

%[Image 2: Fredrick Reines in an undated picture. Reines proved the existence of the neutrino together with Clyde Cowan in 1956. In 1980, he announced that he had found evidence for neutrino oscillations, implying that neutrinos have mass. Photograph by Ed Nano, courtesy of AIP Emilio Segrè Visual Archives.]

The new neutrino picture of the universe was an enormous hit. The 1980 experimental results and its consequences were featured in headlines around the world, from \textit{Nature} to the \textit{New York Times}, and \textit{Discovery Magazine} to \textit{The Christian Science Monitor}. “Massive Neutrinos: Masters of the Universe?,” a headline in \textit{Science} questioned.\cite{Waldrop1981MassiveUniverse} “From Russia with Mass,” read \textit{Science News}.\cite{1980FromNeutrinos}  “The wily neutrino tries to tame the expanding universe,” famed science fiction author Isaac Asimov remarked when lauding the new neutrino-based cosmological model.\cite{Asimov1981TheMouse} Many newspapers featured quotes by Reines: “The consequences would mean man’s view of the universe and his place in it would be profoundly altered.”\cite{1980GivingNeutrinos} Indeed, physicists were just as quick to respond to the results. From recent Nobel-laureate Sheldon Glashow to Soviet cosmology guru Yakov Zeldovich, scientists started to calculate the cosmic consequences of the massive neutrino. It became \textit{the} hot topic in both cosmology and particle physics of the early 1980s. Alex Szalay, after being on the verge of quitting physics to take up a career in music, remembers he “suddenly got two postdoctoral offers” (A. Szalay, oral history interview with J. de Swart, 12 October 2019).

The appeal of the “neutrino-dominated universe” was its capacity to offer a complete picture of the cosmos, from the big bang to galaxy formation. Starting from the homogeneous hot soup of the early universe, the low-mass and weakly-interacting neutrinos would be the first particles to decouple from the mixture of primordial particles, forming gravitational wells long before ordinary matter could. Primordial massive neutrinos provided the seeds for structure formation and give a characteristic mass-scale that governed gravitational collapse. It would also explain why the temperature of the cosmic microwave background was observed to be so uniform:\cite{Partridge1980NewBackground} in this scenario, there is no clumpy distribution of baryonic mass in the early universe that could have disturbed the cosmic microwave background. With massive neutrinos, cosmological structure formed through what was called the “pancake” theory of Yakov Zeldovich: large clusters formed first, and smaller galaxies followed. The Russian team was excited. “Cosmology with massive neutrinos is so attractive that, to paraphrase Voltaire, if it did not exist, it would be necessary to invent it,” a paper by Yakov Zeldovich, Alex Szalay, and collaborators read.\cite[p.32]{Doroshkevich1981CosmologicalMass} There was good reason for the excitement. For the first time, a complete and consistent particle-based picture of cosmic evolution emerged: it explained structure formation, the missing mass, a flat universe, and was supported by grand unified theories. All with the wily neutrino.

\begin{figure}
    \centering
    \includegraphics[width=0.7\linewidth]{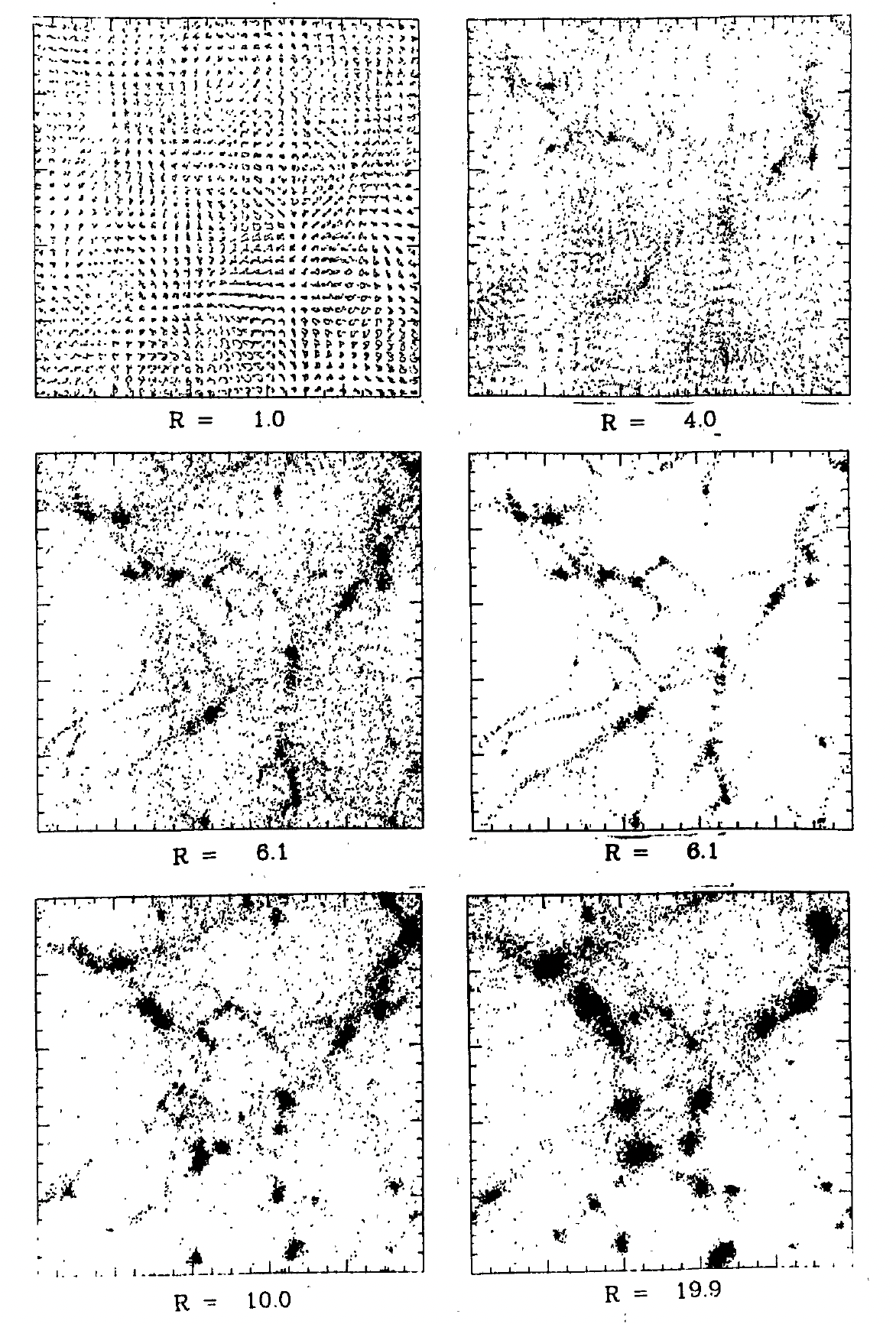}
    \caption{Numerical simulations of the gravitational clustering of mass in a neutrino-dominated universe by Carlos Frenk, Simon White, and Marc Davis, presented in November 1983. \cite{Frenk1984ElementaryContributions} The panels show the mass clustering at different expansion factors, R. The right-hand panel at expansion factor R=6.1 shows the expected distribution of galaxies. The dense clustering in these simulations was seen as strong evidence against the validity of a neutrino-dominated cosmological model. As the authors concluded: "[a] glance at any map of the galaxy distribution on the sky will convince the reader that galaxies are nowhere nearly as strongly clustered" \cite[p. 261]{Frenk1984ElementaryContributions}. Image reproduced from \cite[p. 262]{Frenk1984ElementaryContributions}, ESO/CERN.} 
    \label{fig:frenk}
\end{figure}

The neutrino-dominated dream, however beautiful it was, was short-lived. At the University of California, Berkeley, a crew had gathered around computational cosmologist, and former student of Jim Peebles, Marc Davis. The goal of Davis and team was to simulate the growth of structure in the universe with a computer model. With postdocs Simon White and Carlos Frenk---and equipped with a room-sized VAX 11/780 with 16 Mb of memory---Davis simulated the gravitational clustering of 30,000 galaxies and checked the outcome with recent observations of galaxy clustering (CfA 1 Redshift Survey from 1982). It was among the first cosmological simulations of how cosmic structure would grow from initial fluctuations (see Figure \ref{fig:frenk}). However, running the simulation with neutrinos gave the wrong answer: neutrinos caused much stronger large scale clustering than the observations showed---the relativistic neutrinos smoothed out any small initial gravitational instabilities, causing galaxies to form too late and leaving the universe dominated by large clusters. “The conventional neutrino-dominated picture appears to be ruled out,” the team concluded.\cite[p. L1]{White1983ClusteringUniverse} All the while, the experimental evidence for neutrino mass was slowly evaporating. Richard Feynman had sent around a fierce response to Reines’ results, pointing to faulty analysis.\cite{Vogel2003Interview2003}  “Too bad this talk was ever given,” Reines lamented in his personal notes about the announcement of his neutrino oscillation results.\cite{FrederickCalifornia} 
%neutrino density contrast was too high comapred to observations. Observed galaxies are considerably less clustered than in the neutrino simulation.  

%[Image 3: Neutrino-dominated universe. TBD.]

\section*{WIMPs and the norms of particle cosmology}

Neutrino dark matter had become unfashionable by 1983, but the idea that dark matter is a particle was there to stay. The success of the neutrino could be emulated with a different particle and solved the neutrino’s issue with galaxy formation. Peebles and others found that, instead of the particle being \textit{hot} in the early universe---that is moving at relativistic speeds like the neutrino, washing out any small-scale gravitational instabilities---the dark matter particle needed to be \textit{cold}: heavier and slower moving particles allowed for small initial instabilities and galaxies, not clusters, to form first.\cite{Peebles1982,Blumenthal1982GalaxyNeutrinos} Although Peebles “didn’t mean it to be taken seriously,” (P. J. E Peebles, oral history interview with J. de Swart, 13 November 2014) the cold dark matter scenario was immediately widely embraced. It was “the best model available,” an influential 1984 \textit{Nature} review read: cold dark matter predicted the right spectrum of initial density fluctuations, masses for galaxies, haloes of dark matter, and the uniformity of the cosmic micro-wave background—all matching the observations.\cite{Blumenthal1984} Simulations by Marc Davis and team swiftly confirmed its success.\cite{Davis1985}

Unsuspectingly, these cold dark matter particles were familiar to many particle physicists: the theorized “heavy neutrinos” introduced \textit{en masse} in 1977. By the early 1980s, several new physics proposals included hypothesized particles that fitted this broad weakly-interacting category: the theory of super symmetry predicted particles like the \textit{gravitino} that could be the dark matter, Heinz Pagels and Joel Primack argued;\cite{Pagels1982} or it might be the \textit{photino}, a CERN-Stanford team suggested.\cite{Ellis1984SupersymmetricBang} There was no shortage of other hypothetical particles suggested by theoreticians: Pierre Sikivie forwarded the \textit{axion} particle as cold dark matter\cite{Ipser1983CanAxions}---the product of a solution to a symmetry problem in the standard model---and Edward Witten showed that dark matter might be invisible \textit{nuggets of quarks} leftover from early universe phase transitions.\cite{Witten1984CosmicPhases} \textit{Cosmions}, some called the particles in this “zoo” of dark matter candidates—that is, until 1985, when Gary Steigman and Michael Turner dubbed the popular electro-weak-scale candidates WIMPs, \textit{Weakly Interacting Massive Particles}.\cite{Steigman1985CosmologicalParticles}

WIMPs were borne from the ashes of the neutrino-dominated universe, and so did new standards for using particle physics in cosmology. Where in the 1970s cosmology was used to \textit{constrain} particle physics theories, the neutrino scenario had shown that particle physics was indispensable to \textit{explain} the cosmos and its large-scale features: structure formation, the smoothness of the CMB, the flat universe, and the missing mass. The new interdisciplinary field of ‘particle cosmology’ flourished: dedicated meetings on particles and the universe were organized, and at Fermilab the first particle astrophysics department was created—the first NASA-funded program at a U.S. national lab. Particle physics became the norm through which cosmological theories were to be evaluated. It is hence not surprising that, when in 1983 Mordehai Milgrom introduced a different explanation for dark matter based on an alternative force law,\cite{Milgrom1983AHypothesis.} the response was skeptical. Aside from failing to address cosmic structure formation, “[i]t is not at all clear what sort of particle physics could lead to a force law like [Milgrom’s],” Joel Primack contended in 1984.\cite[p. 63]{Primack1984DarkUniverse} 

Not only did the neutrino universe urge new theoretical norms, it also provided an experimental backbone to uncover the WIMP. In the wake of the 1980 neutrino-detection craze, Munich physicists Andrzej Drukier and Leo Stodolsky had introduced a new detector principle to measure mass and properties of neutrinos.\cite{Drukier1984PrinciplesAstronomy} When their paper landed on the desk of theoretical physicist Edward Witten at the Institute of Advanced Study in Princeton, Witten argued that the same technology might be used to detect the newly celebrated WIMPs.\cite{Goodman1985DetectabilityCandidates} A team at Harvard, including the recently relocated Drukier, realized that this WIMP-parameter space was within reach of existing neutrino experiments.\cite{Drukier1986DetectingCandidates} Working with nuclear physicists at a neutrino experiment at Pacific North-West Laboratory (PNL), famed for its enrichment of plutonium and later nuclear non-proliferation interests, they produced the first WIMP direct detection results. The PNL experiment set limits on the mass of the WIMP that were unmatched for at least a decade—and inspired a WIMP detection infrastructure that endures to this day.\cite{Ahlen1987LimitsSpectrometer}

WIMP detection efforts continue. And while cosmic neutrinos are threatening the experimental future and theoretical promise of the WIMP, they are deeply entangled with the roots of the paradigm: they were the first dark matter particle candidates, legitimized the field of particle cosmology, and even provided the first experimental apparatuses. As a gateway to the WIMP paradigm, they provided a precursor to what was later called the “WIMP miracle”: the existence of a particle whose amount and interaction strength creates the perfect picture of density fluctuations and the evolution of structure in the universe. Most of all, the neutrino-dominated universe showed the massive explanatory potential of a particle dark matter picture, integrating galaxy formation, large-scale structure, nucleosynthesis, GUTs, the cosmic microwave background, primordial density fluctuations, and even inflation. Although interest in the WIMP is waning today, it is thus important to realize that the “WIMP paradigm” is not just about the potential existence of a new particle. Paradigms also comprise norms about what is considered good science, which questions are worth addressing, and what experimental instrumentation is deemed proper.\cite{Kuhn1962} The WIMP paradigm of dark matter includes deeply rooted particle physics standards and explanatory expectations in which the success of the neutrino-dominated universe reverberates. Cosmic neutrinos will have to work hard to erase their own legacy.

\begin{framed}

\section*{Box 1: The Neutrino Fog and the Limits of WIMP Detection}
Since the first experimental efforts in the 1980s, it has been clear that detecting dark matter in the form of hypothetical particles interacting through the electro-weak force would critically depend on reducing backgrounds. Aimed at detecting the recoils of these \textit{Weakly Interacting Massive Particles }(WIMPs) on target nuclei, these experiments have progressively minimized sources that could mimic WIMP signals. This has meant deploying experiments in deep underground laboratories, constructing them from ultra-radiopure materials, and building sophisticated filtration systems to remove radioactive contaminants.\cite{deSwart2025CleaningElusiveness} Continued null results have driven the field to ever-lower backgrounds and larger detector volumes---improving sensitivity by about eight orders of magnitude over the past four decades.\cite{Gaitskell2004DirectMatter} %And making the WIMP currently around x times the proton (1 GeV/c2 - 100 TeV/c2),

Yet, with this extraordinary sensitivity, a previously negligible background has become pertinent: neutrinos. Neutrinos from the Sun, atmosphere, and supernovae can scatter off the target nuclei via coherent elastic neutrino-nucleus scattering ($CE\nu NS$), producing nuclear recoils that are virtually indistinguishable from WIMP interactions.\cite{Monroe2007NeutrinoSearches} Each neutrino source mimics WIMPs of different masses and cross sections. Because neutrinos cannot be shielded, the onset of this background was dubbed the neutrino floor.\cite{Grothaus2014DirectionalBound} As sensitivities approached this regime, the terminology changed: not a floor but a “neutrino fog”, gradually obscuring WIMP signals depending on neutrino fluxes and detector statistics.\cite{OHare2021NewSearches}

By the end of 2025, the XENONnT, PandaX, and LZ collaborations had reported the first indications of this fog.\cite{Aprile2024FirstXENONnT,PandaXCollaboration2024FirstPandaX-4T,Akerib2025SearchesExperiment} While a major milestone, these results also push the hypothetical particle into previously unimagined parameter spaces that are possibly unreachable with known technologies. The neutrino fog thus presents an existential challenge---not only to the detection of WIMPS, but to the paradigm itself. While a major collaboration is now planning to build what may be the “ultimate” WIMP experiment,\cite{Aalbers2016DARWIN:Detector}  many in the field are shifting focus, broadening the search for alternative dark matter candidates. The risk remains that, if they exist, WIMPs may be forever hidden in the irreducible flux of astrophysical neutrinos---the particles that, ironically, inspired the efforts to detect WIMPs in the first place. 
    \label{fig:box}

\end{framed}

\bibliography{deSwart_Neutrino.bib}

\section*{Competing interests}
The author declares to have no competing interests. 

\section*{Acknowledgments}
The author gratefully acknowledges his interviewees who were willing to take time to share their recollections, the archivists at UC Irvine, and the American Institute of Physics for awarding him a Robert H. G. Helleman Memorial Postdoctoral Fellowship to be at MIT. He is also indebted to David Kaiser, Ben Lehmann, Marianne Moore, and colleagues at MIT's Center for Theoretical Physics for helpful comments and conversations.

\end{document}